# Compact simultaneous label-free autofluorescence multi-harmonic (SLAM) microscopy for user-friendly photodamage-monitored imaging

GENG WANG,[1] STEPHEN A. BOPPART,[1,2,3,4,5,6] AND HAOHUA TU [1,2,*]

[1] *Beckman Institute for Advanced Science and Technology, University of Illinois at Urbana-Champaign, Urbana, Illinois 61801, USA*

[2] *Department of Electrical and Computer Engineering, University of Illinois at Urbana-Champaign, Urbana, IL 61801, USA*

[3] *Department of Bioengineering, University of Illinois at Urbana-Champaign, Urbana, IL 61801, USA*

[4] *Cancer Center at Illinois, Urbana, IL 61801, USA*

[5] *Carle Illinois College of Medicine, University of Illinois at Urbana-Champaign, Urbana, IL 61801, USA*

[6] *Center for Label-free Imaging and Multi-scale Biophotonics (CLIMB), Urbana, IL 61801, USA*

\* *htu@illinois.edu*

**Abstract:** Label-free nonlinear optical microscopy has become a powerful tool for biomedical research. However, the possible photodamage risk hinder further clinical applications. To reduce these adverse effects, we constructed a new platform of simultaneous label-free autofluorescence multiharmonic (SLAM) microscopy, featuring 5-channel multimodal imaging, inline photodamage monitoring, and pulse repetition-rate tuning. By the use of a birefringent photonic crystal fiber for spectral broadening (rather than supercontinuum generation) and a prism compressor for pulse pre-chirping, this system allows users to independently adjust pulse width, repetition rate, and energy, which is useful for optimizing imaging condition towards no/minimal photodamage. Also, it demonstrates label-free multichannel imaging at one excitation pulse per image pixel and thus paves the way for improving the imaging speed by a faster optical scanner.

## 1. Introduction

Intravital imaging technology with high speed, sufficient spatial resolution, and long-term photodamage-free capability is critical to the study of biological processes. It permits direct and longitudinal tracking of diverse intercellular behaviors in their native environment instead of inferring possible processes based on static images [1-4]. As one type of intravital imaging, label-free nonlinear optical microscopy [5-8] has become a powerful tool for biomedical research due to its advantages of low invasiveness, deep imaging penetration, high resolution, etc., especially in neuroscience, oncology and immunology [9-12]. However, the relatively slow imaging speed and the accompanying photodamage risk bring limitations to further preclinical and clinical studies. The improvement of imaging speed and the suppression of photodamage can be achieved by different methods, such as improving the excitation efficiency [7,13], using a polygonal mirror and resonant scanner [14], splitting one pulse to sub-pulses with equal energy [15], applying an adaptive light source to illuminate only the area of interest (ROI) [16], or improving the signal-to-noise ratio(SNR) of a single frame through a deep learning algorithm [17,18]. One fundamental factor that limits the imaging speed is the low excitation efficiency of a single pulse. For a typical 80 MHz imaging system [19], the power of each laser pulse has to be kept low in order to avoid heating-related photodamage. Therefore, in order to obtain an image with acceptable SNR, each pixel needs to contain tens to hundreds of pulses, which greatly limit the imaging speed.

In 2018, by a combination of reduced repetition rate of the laser source (10 MHz) and shortened pulse width (<60 fs) at 1110-nm central wavelength, You et al. obtained a relatively high peak power of optical pulses and demonstrated simultaneous label-free autofluorescence multi-harmonic (SLAM) microscopy [7], which can simultaneously acquire the signals of two- and three-photon excited autofluorescence (2PAF/3PAF) of FAD/NAD(P)H, respectively, and second/third harmonic generation (SHG/THG), by using the supercontinuum generation technology based on a photonic crystal fiber (PCF). However, several limitations remained: 1) The high peak power supercontinuum generation leads to a short life of the PCF (~200 hr), which needs to be replaced regularly and thus complicates the operation of the system; 2) The limited average power for a given excitation band (~60 nm) does not allow users to adjust pulse repetition rate in a wide range to meet different imaging requirements; 3) Although the peak power of excitation pulse is greatly improved, the photon-counting photomultiplier (PMT) detection requires a large number (>10, typically) of excitation pulses for each pixel (a long pixel dwell time) for sufficient signal (counts), which limits imaging speed; 4) The components used in the system are large, resulting in a rather bulky and complex system.

Here, we demonstrate a new and compact SLAM platform and system. The central wavelength of the excitation window is shifted to ~1030 nm, which is much accessible commercially. By using a PCF and a prism compressor, we obtain excitation pulses from near-transform-limited 60 fs (FWHM) to uncompressed 300 fs with a broad bandwidth (~80 nm (990-1070 nm), bottom-to-bottom), and sufficient pulse energy (or average power) under a wide and tunable repetition rate (800 kHz - 20 MHz). More importantly, these three pulse parameters (width, repetition rate, energy) can be adjusted independently without interference. This allows users to find the optimal imaging conditions for different applications to maximize the signal-to-photodamage ratio. With only PCF-assisted spectral broadening free of supercontinuum generation, the laser source is stable indefinitely (> 1 year) without the need to replace the PCF. Finally, the higher peak power permits single pulse per pixel label-free imaging via analog PMT detection, and can greatly increase the imaging speed.

## 2. Materials and Method

### 2.1 High excitation efficiency with low duty cycle

For label-free nonlinear imaging, it is critical to achieve high excitation efficiency without photodamage. According to the nonlinear optical signal generation formula (*Eq.1*), when the average power $P$ remains unchanged, the signal generation intensity $S$ can be significantly increased by reducing the duty cycle, consisting of laser repetition rate $f$ and pulse width $\tau$, especially for high order nonlinear processes:

$$S \sim (f\tau)\left[\frac{P}{f\tau}\right]^n = \frac{P^n}{(f\tau)^{n-1}} \tag{1}$$

where $n$ denotes the order of nonlinear process (2 for 2PAF/SHG, 3 for 3PAF/THG, 4 for 4PAF). For two/three/four-photon imaging, when the duty cycle is reduced by a factor of 4, the signal will increase exponentially and increase by a factor of 4/ $4^2$/ $4^3$, respectively. Therefore, by reducing the duty cycle, equal or higher signal can be achieved using lower average power (lower risk of thermal damage). More importantly, it can enable the excitation of higher-order nonlinear processes.

## 2.2 Excitation window

For this new platform of SLAM microscopy, the central wavelength was shifted to 1030 nm. Compared with the previous 1110 nm center wavelength, the 1030 nm center wavelength has several advantages: 1) Relatively short excitation wavelength can provide higher multiphoton excitation efficiency for FAD, NAD(P)H and tryptophan; 2) The excitation center wavelength is the same as the source laser emission center wavelength for PCF spectra broadening, so there is no need to use high peak power to generate supercontinuum, and only need to use low peak power to broaden the spectrum in the PCF to achieve a near-transform-limited pulse, which can avoid the problem of PCF damage and replacement; 3) Lower input peak power for PCF allows the same spectral broadening over a wide range of repetition rates (with a wide range of average power) to meet the needs of different applications,.

## 2.3 Fiber source

The laser source for PCF spectral broadening is a compact ultrafast laser (Satsuma, Amplitude Laser) with $1030\pm5$ nm central wavelength (inset a of Fig. 1), 10 W maximum average output power, < 350 fs pulse width, and tunable repetition rate from single shot to 40 MHz. In order to obtain near-transform limited pulses to improve the multiphoton signal generation efficiency, laser pulses were first sent into a 25-μm-core PCF (LMA-25, Thorlabs) to achieve a coherent spectral broadening of around 80 nm ($1030 \pm 40$ nm, inset a of Fig. 1) with ~80% coupling efficiency, and then the collimated beam was sent to a prism pulse compressor (BOA-1050, Swamp Optics) for dispersion compensation to obtain 60 fs near-transform limited pulses, which can be achieved by adjusting the group delay dispersion (GDD) via the compressor, as shown in the inset b of Fig. 1. In practical application, we should pre-chirp the pulse to compensate for the dispersion caused by the optical components after the compressor. Compared with the previous SLAM system [7], the new platform does not need to select an excitation window at 1110 nm in the supercontinuum, so this eliminates the need of the expensive pulse shaper.

## 2.4 System setup

We designed a new SLAM platform and system (Fig.1). The $1030 \pm 40$ nm pulses from the prism compressor were raster scanned by a 4 mm Galvo-Galvo Scanner (LSKGG4, Thorlabs) and focused by an inverted high UV-transmission objective (UAPON 40XW340, N.A. = 1.15, Olympus) with an ample adjustable average power on the sample after the loss along the excitation beam path. One telecentric scan lens (SL50-2P2, *f=50* mm, Thorlabs) and one tube lens (TTL200MP, *f=200*, Thorlabs) expanded the beam size to fill the back focal plane pupil of the objective (Ø10.35mm). Since high power is not required to generate supercontinuum, this system with its PCF can operate over a wide range of repetition rates (800 KHz~20 MHz) with similar spectra broadening, as shown in the inset a of Fig. 1 (1/2.5/10 MHz). For example, when the repetition rate is 2.5 MHz, the excitation power before PCF is only 180 mW, but the maximum average power on the sample can exceed 50 mW, which gives users a high power dynamic range for different applications. Based on the objective lens we use, the typical maximum field of view (FOV) is around 400 μm× 400 μm, the actual FOV depends on the scanning angle of the galvanometer-driven mirror.

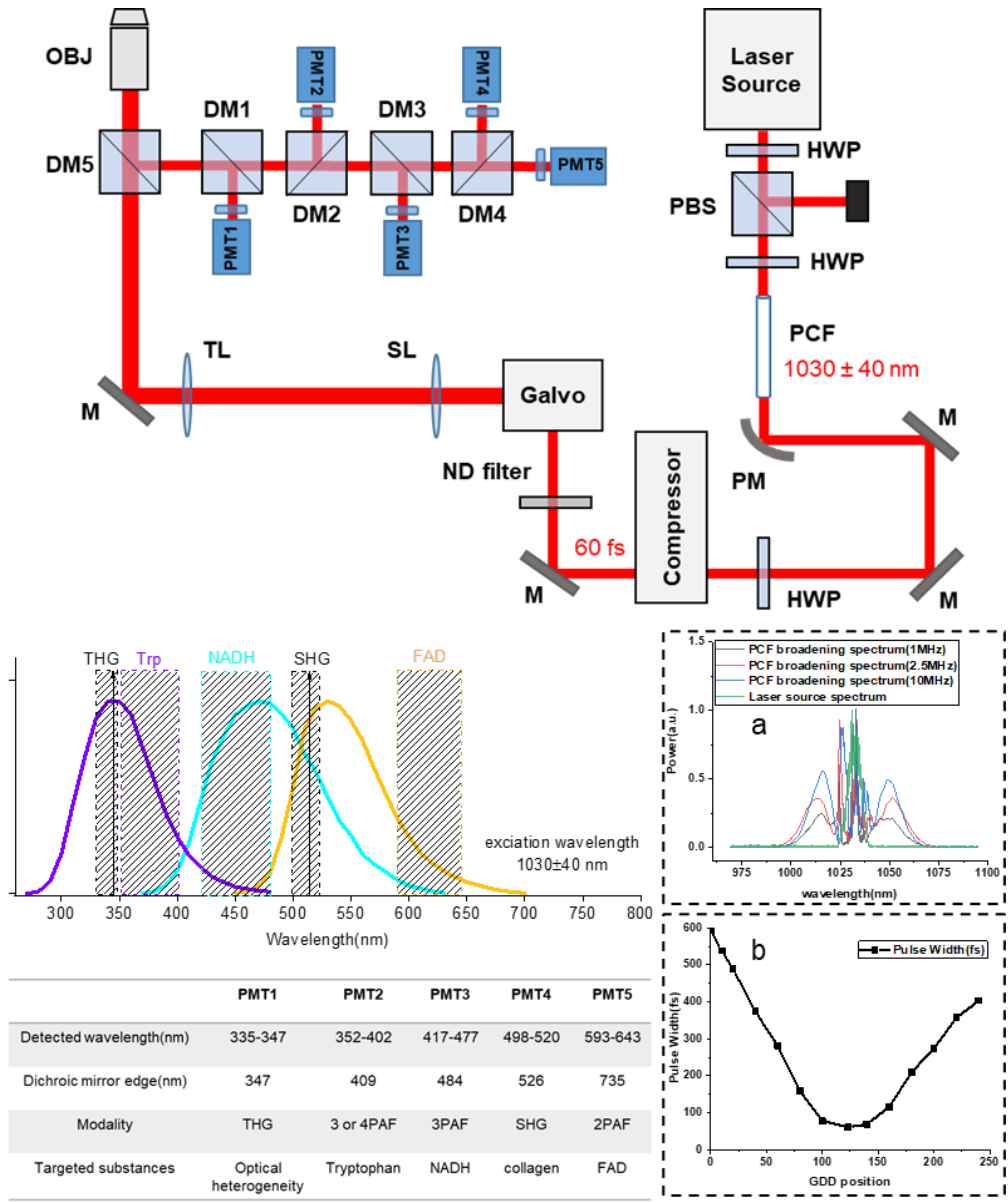

Fig.1 Schematic of the compact SLAM system with five spectral detection channels and the emission spectra of tryptophan, NAD(P)H and FAD. High peak power pulses were sent into the PCF to broaden the spectrum. Prism compressor was used to compensate the dispersion to obtain near-transform limited pulses. Multimodal multiphoton signals were spectrally- separated into 5 detection channels by long-pass dichroic mirrors and specific bandwidth filters and detected by PMTs as listed in the table. Inset a: spectra before (green) and after (black- 1 MHz, red- 2.5 MHz, blue-10 MHz) spectrum broadening based on birefringent PCF. Inset b: Measured results of compressor GDD position and pulse width (for 2.5 MHz/ 180 mW input). HWP half wave plate, PBS polarizing beamsplitter, PCF photonic crystal fiber, PM parabolic mirror, M mirror, ND neutral density, SL scan lens, TL tube lens, DM dichroic mirror, OBJ objective, PMT photomultiplier tube.

## 2.5 Channels for detection

Regarding the detection part, this system is designed to detect NAD(P)H from the 3PAF channel, FAD from the 2PAF channel, and tryptophan from the mixed 3PAF/4PAF(potential) channel, combined with noncentrosymmetric structures from the SHG channel and interfacial features from the THG channel, all simultaneously. The reflected multimodal multiphoton signals were spectrally-separated into 5 detection channels by long-pass dichroic mirrors and appropriate band-pass filters, as shown in Fig. 1. To minimize the crosstalk between individual channels and maximize the detection efficiency of each channel, the 5 channel filters were chosen with specific bandwidth characteristics and listed in Fig. 1. For this system, the high pulse peak power can enable more than one photon excitation per pulse when operating at low repetition rates. Therefore, the spectrally-separated signals were detected by analog PMTs instead of photon-counting PMTs (PMT2101, Thorlabs for 3PAF/ mixed 3PAF and 4PAF(potential)/ 2PAF/ THG channels; PMT1001, Thorlabs for SHG channel).

*2.6 Imaging acquisition*

To make the operation of this system more user-friendly, a commercial control software (Scanimage, Vidrio Technologies, LLC) is used, which has powerful features and can allow users to set different scanning schemes by customizing key parameters, such as the number of pulses per pixel, FOV, repetition rate, etc. The details of the configuration control interface can be found in reference [20]. For this compact SLAM platform, since the sampling clock is synchronized with the fundamental clock of laser source, the number of pulses per pixel can be set via the pixel bin factor (single/multi- pulse), which is limited by the line period (cannot exceed the maximum scanning rate of the galvo scanner) and also related to image pixel size. The frame rate is related to line period and image pixel size, and the pixel dwell time is set by a combination of the repetition rate and pixel bin factor. Furthermore, since a pulse picker is used in this laser to adjust the repetition rate, the tunable repetition rate should be divisible by the fundamental rate of laser source (40 MHz). When the repetition rate is 0.83 MHz and the image size is 1024×1024 pixels, one pulse per pixel can be set at a frame rate of 0.7 Hz. Similarly, a minimum of 2/3/6/12/24 pulses per pixel can be set at 1.67/2.5/5/10/20 MHz repetition rate, and the number of pulses per pixel can be increased by reducing the scanning speed of the galvo scanner. In summary, the frame rate, line rate, pixel dwell time and average power on sample are dependent parameters. Once the repetition rate, pulse energy, bin factor and image pixel size are fixed, these dependent parameters are also determined. The main independent parameters of this system are shown in Table 1.

Table 1. Independent Parameters of the compact SLAM

| typical repetition rates (MHz) | 0.83, 1.67, 2.5, 5, 10, 20 |
|---|---|
| typical pulse width (fs) | 60~300 |
| pulse energy (nJ) or average power (mW) on sample | <5 or <20 (limited by photodamage) |
| central wavelength (nm) | ~1030 (fixed) |
| frame size (pixel) | 1024×1024 (fixed) |
| number of pulses per pixel (bin factor) | 1- 24 pulses |
| data acquisition along fast scanning direction | bidirectional |
| zoom factor (or pixel size) | diffraction-limited sampling (or fixed FOV of 400 μm× 400 μm) |
| objective (magnification, NA) | 40×, 1.15 (fixed) |
| imaging depth (μm) | <200 (limited by working distance of the objective) |

Altogether, by using a 25 μm core PCF and prism pulse compressor, we obtained excitation pulses from near-transform-limited 60 fs to uncompressed 300 fs with broad bandwidth (~80 nm), and sufficient average power (or pulse energy) under a wide and tunable repetition rate (800 kHz - 20 MHz). More importantly, these three parameters can be adjusted independently without interference. This allows users to find the optimal imaging conditions for different applications.

## 3. Results

*3.1 Ex vivo mouse tissue imaging*

Label-free mouse tissue was used to demonstrate the imaging capabilities of this system, as shown in Fig. 2. Among them, Fig. 2a-2e were acquired at a repetition rate of 2.5 MHz with 10 mW average power on the sample. The FOV is around 400 μm×400 μm with 1024×1024 pixels per frame and 3 pulses per pixel in one single frame. The images shown were obtained by averaging 10 frames with 0.7 Hz single frame rate (1024×1024 pixels per frame). It is worth mentioning that the current frame rate is limited by the relatively slow scan rate of the galvo-driven mirror (500 Hz maximum), which can be addressed by using a high-speed rotating polygonal mirror or a resonant scanner. The images within the dashed box were the fat-rich areas on the inner surface of the mouse skin (Fig. 2a-2e). The rich information of 4 different modalities is clearly reflected through the 4 different spectral channels. Among them, channel 1 is the 3PAF signal reflecting NAD(P)H information (Fig. 2b), channel 2 is the THG signal (Fig. 2c), channel 3 is the SHG signal (Fig. 2d), and channel 4 is the 2PAF signal reflecting FAD information (Fig. 2e). Clear adipocytes and cell membranes are reflected in the THG channel, and abundant fibers are reflected in SHG channel. The composite image with 4 different modalities is shown in Fig. 2a.

The label-free multi-channel imaging capability with a single excitation pulse per image pixel was also demonstrated through the sample of *ex vivo* mouse kidney, as shown in Fig. 2f and 2g, which were collected at a repetition rate of 0.83 MHz with 3.6 mW power on sample and comprise the information of 4 channels (2PAF, 3PAF, SHG and THG). The FOV is around 250 μm×250 μm with 1024×1024 pixels per frame. Fig. 2f is a single-frame imaging that contains one pulse per pixel, and Fig. 2g (20 pulses per pixel) is the result of averaging 20 frames under the imaging conditions of Fig. 2f. Comparing the two images, it can be seen that the single-frame imaging with a single pulse per pixel can reveal the tissue structure with a relatively low SNR. If paired with a high-speed scanner, the system could achieve high-frame-rate, label-free, long-term imaging with a low risk of nonlinear photodamage.

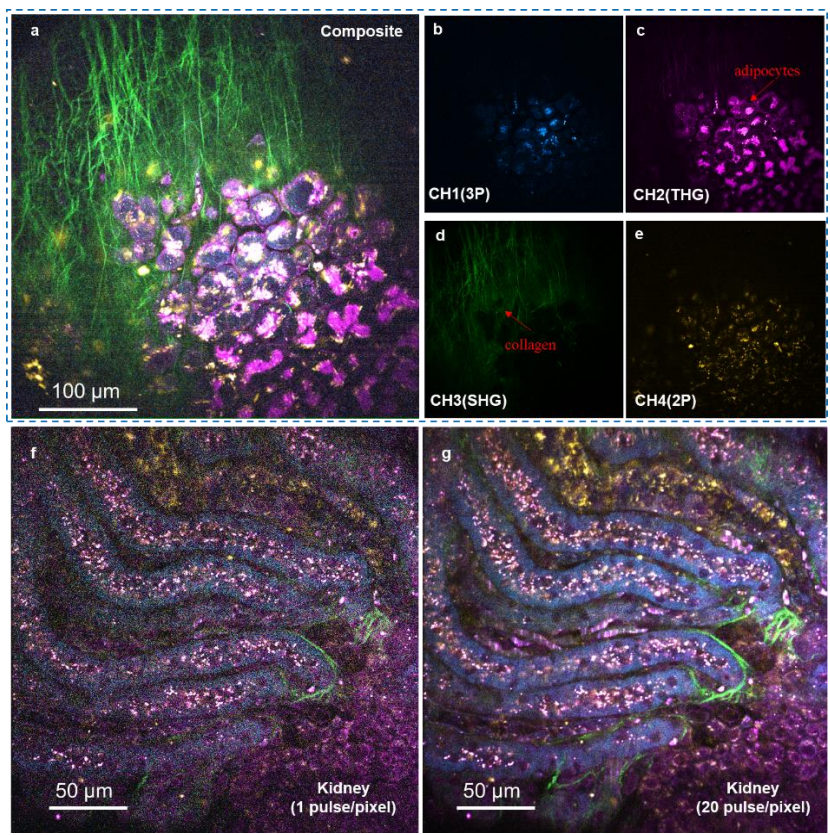

Fig.2 *Ex vivo* label-free imaging of mouse tissue by the compact SLAM system. Color in this figure: cyan (3PAF), magenta (THG), green (SHG), yellow (2PAF). (a) imaging of the inner surface of the mouse skin with 4-channels composite information.

(b) 3PAF signals with NAD(P)H information, (c) adipocytes revealed in THG signals. (d) collagen fibers revealed in SHG signals, (e) 2PAF signals with FAD information, (f) single-frame imaging of a mouse kidney (composite) at one excitation pulse per pixel. (g) image averaged over 20 frames under the same imaging condition of (f).

*3.2 Photo-damage monitored imaging*

Photodamage is a common problem in the field of optical bioimaging and is considered throughout different imaging techniques [21-24]. This compact SLAM system with tunable ultrashort pulses can generate enough power on the sample to allow users to find the optimal imaging conditions for different applications based on monitored-photodamage limits and SNR.

Due to the high spatial homogeneity, *ex vivo* chicken breast was used in experimental studies of photodamage. By adjusting the power on the sample, we observed the photodamage of the sample in the time-lapse imaging. For the homogeneous chicken breast tissue, almost no autofluorescence signal was observed in the 3-photon channel without photodamage. However, we can observe the structure of muscle tissue through the SHG channel for focusing and determining the ROI, as shown in Fig. 3a.

We found that the 3PAF signal does not change in fluorescence intensity over time (200 frames in total, 285 seconds) when the power was low (9 mW on sample/ 2.5 MHz/ 60 fs), as shown in Fig. 3b (blue line). With the increase of the power on sample, some newly generated fluorescent signals were observed in the 3PAF channel first and increased over time (200 frames in total, 285 seconds, Fig. 3b red line), as shown in Fig. 3d~ 3f (20 mW on sample/ 2.5 MHz/ 60 fs), which were caused by the emergence of new substances due to photodamage [25,26]. The growth rate of fluorescence intensity in the 3PAF channel increases with the increase of power over time until bubbles were generated (Fig. 3f). This phenomenon can be observed not only in muscle tissue, but also in cells, as shown in Fig. 3c. Because different samples have different damage thresholds, by changing the power, repetition rate and pulse duration, the users can find optimal imaging conditions to maximize the signal-to-photodamage ratio for their applications.

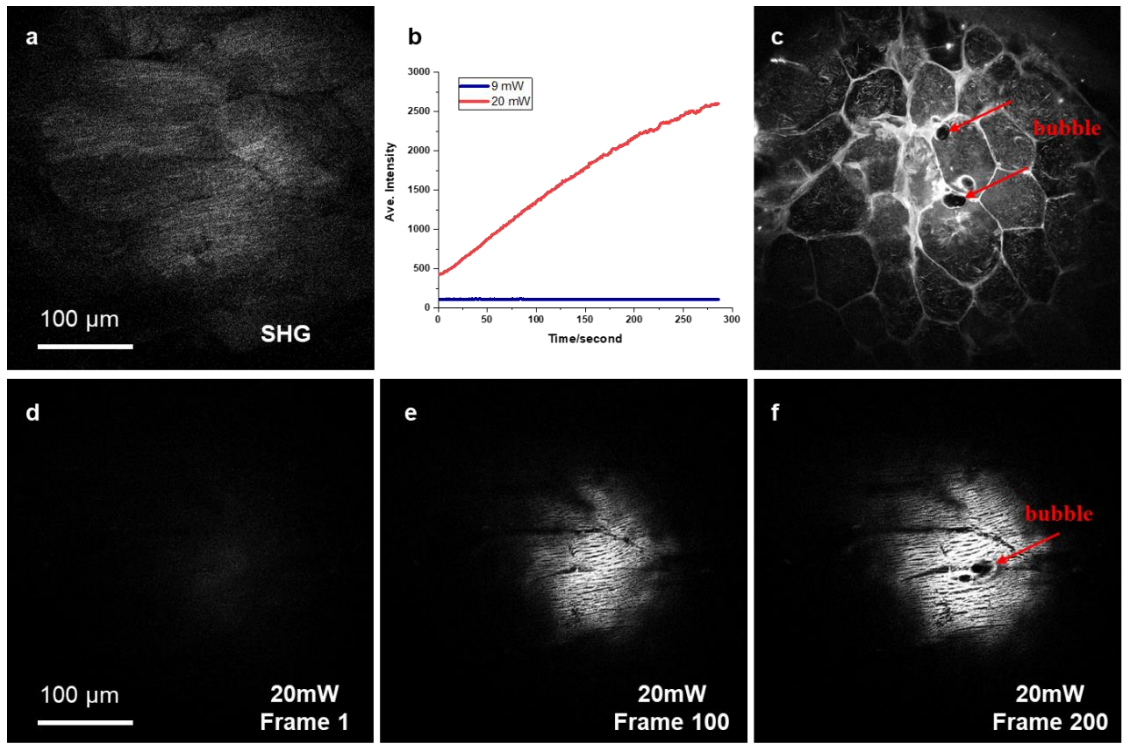

Fig. 3 Photo-damage monitored imaging in tissue and cells by the compact SLAM. (a) SHG channel was used to observe the muscle structure of the chicken breast to find the ROI for photo-damage experiment. (b) Changes in the average intensity of 3-

photon channel images of chicken breast over time (d-f) under irradiation of 9 mW and 20 mW laser power. (c) Photodamage observed trough 3-photon signals of bacon fat cells with the generation of clear bubbles (2.5 MHz / 60 fs). (d)~(f) The 1st, 100th, and 200th three-photon channel images of chicken breast under 20 mW irradiation, which correspond to the red line in Fig. 4b.

*3.3 Time-gated window for denoising*

For laser scanning fluorescence imaging, the signal occupies only a small part between laser pulses, and the rest is useless noise, especially for the low repetition rate laser imaging system, as shown in Fig. 4a. Since each pixel represents the average fluorescence signal integrated over the dwell time of that pixel, integrating these noisy regions within the signal can greatly reduce the SNR of the image, especially for the weak signal imaging. To reduce this adverse effect, the system applied a time-gated sampling window for each laser pulse through the FPGA to extract the useful fluorescence signal. An *ex vivo* mouse kidney sample was imaged to illustrate the effect of applying time-gated windows with different durations, as shown in Fig. 4b (10 ns) and 4c (100 ns), and the time interval between two laser pulses was 200 ns (5 MHz repetition rate) for this test. Compared with the image with the shorter duration sampling window (Fig. 4b), the scheme for using a longer duration sampling window (Fig. 4c) or integrating over the whole region greatly reduced the SNR and detection limit of the image, resulting in weak signal regions being buried in the noise, as shown the areas indicated by red arrows in Fig. 4b and 4c. As for the duration setting of the sampling window, it should be longer than the fluorescence lifetime of the sample.

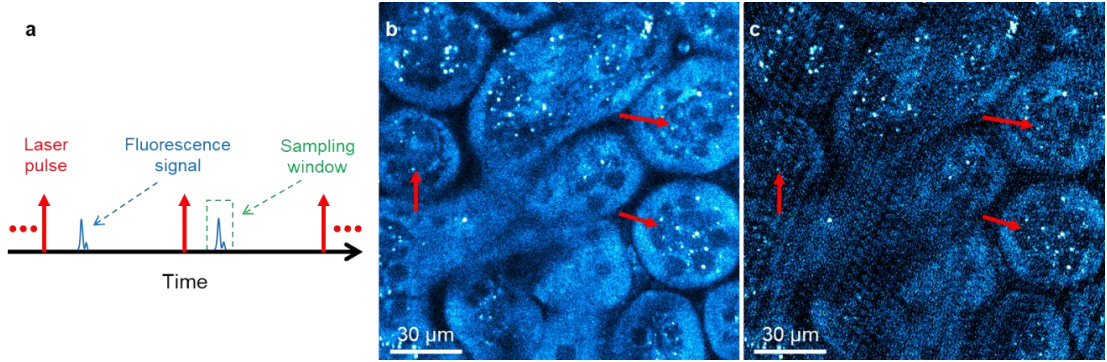

Fig. 4 Time-gated window for noise reduction. (a) Schematic of time-gated window. (b) Three-photon mouse kidney image with a 10 ns sampling window for 5 MHz system. (c) Three-photon mouse kidney image with a 100 ns sampling window for 5 MHz system.

*3.4 Power/concentration dependent experiments*

To verify the multiphoton excitation performance of this compact SLAM system, power-dependent experiments were performed. If we conduct logarithm operation on both sides of *Eq. 1*, we obtain:

$$\log_{10} S = n \cdot \log_{10} P + (1-n) \cdot \log_{10}(f\tau) \qquad (2)$$

where the second term on the right side of the *Eq. 2* is a constant. If $(1-n)\cdot\log_{10}(f\tau)$ is replaced by $c$, $\log_{10} S$ is replaced by $y$, and $n\cdot\log_{10} P$ is replaced by $x$, *Eq.2* can be simplified to *Eq.3* :

$$y = n \cdot x + c \qquad (3)$$

Therefore, through power dependent experiments and linear fitting, the nonlinear order- $n$ of each channel can be measured. For this system, to perform power dependent experiments, an FAD solution was used for the 2PAF channel, an NADH solution was used for 3PAF channel, and the surface signal of a coverslip was used for the THG channel, as shown in Fig. 5a- 5c. The experimental results showed that the linear fitting slope of the two-photon channel is ~2(2.017), and the linear fitting slope of the three-photon and THG channels is ~3(2.984 and 3.027), which is consistent with the theory. Meanwhile, solution concentration dependent experiments were also performed to validate the linear correlation between the concentration and intensity based on an analog PMT, as shown in fig. 5d. The results indicated that solution concentration and signal intensity exhibited a proportional linear relationship without signal saturation. Furthermore, based on the proportional linear correlation between the signal intensity and concentration of autofluorophore, the concentration of multiphoton signal can be quantified based on the calibrated power/concentration dependent experiments, even for the analog PMT. Finally, the above two types of experimental results further support the effective suppression of noise by the time-gated window.

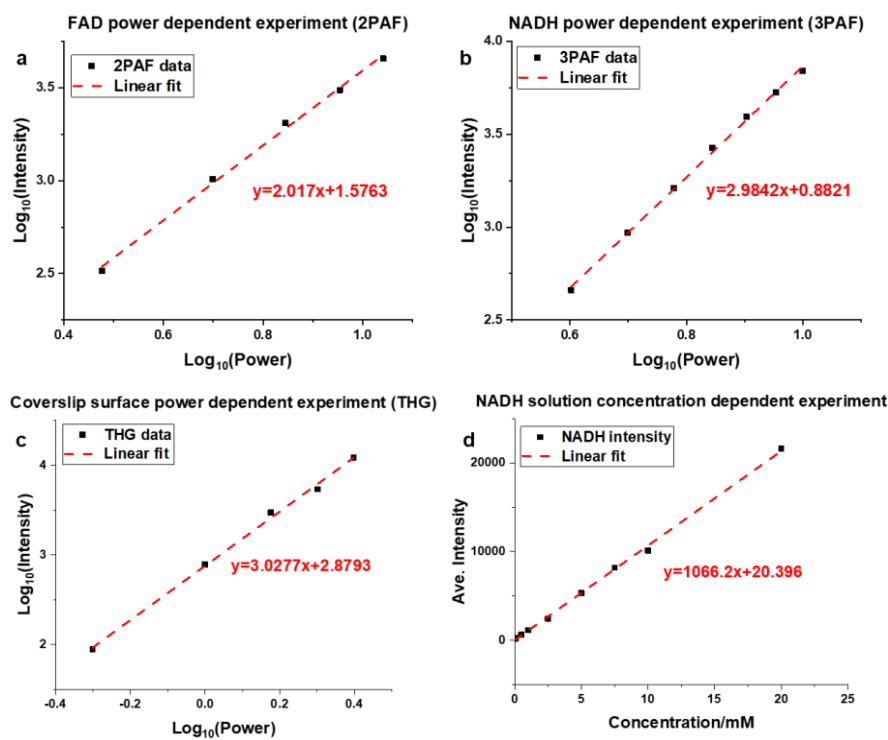

Fig. 5 Power and concentration dependent experiments on the compact SLAM system. (a) 2-photon channel power dependent experiment (3,5,7,9,11 mW on sample) based on 10 mM FAD solution. (b) 3-photon channel power dependent experiment (4/5/6/7/8/9/10 mW on sample) based on 10 mM NADH solution. (c) THG channel power dependent experiment (0.5/1/1.5/2/2.5 mW on sample) based on coverslip surface. (d) 3-photon channel concentration dependent experiment (0/0.1/0.5/1/2.5/5/7.5/10/20 mM) of NADH solution with 4 mW on sample.

## 4. Conclusion

In this paper, we built a compact SLAM system based on off-the-shelf optical and mechanical components, which enabled simultaneous excitation and acquisition of 2PAF/3PAF from FAD/NAD(P)H and SHG/THG signals via a single-excitation source. The high excitation efficiency enabled by the combination of more efficient excitation wavelength and higher peak power improved the SNR of single frame, reduced the requirement for single-pixel dwell time or the total average number of frames, thereby reducing the practical total imaging time and increasing the imaging speed. By using a robust 25 μm large-core PCF and prism compressor, we obtained excitation pulses from near-transform-limited 60 fs to uncompressed 300 fs with broad bandwidth (~80 nm), and sufficient average power under a wide and tunable repetition rate (800 kHz - 20 MHz), and realized one pixel per pulse label-free imaging. Moreover, pulse repetition rate, pulse energy, and pulse duration on sample can be adjusted independently without interference to satisfy different applications. Daily operation of this imaging system has been reliable, as imaging can be routinely performed 10 min after the laser is turned on. The occasional drift in fiber coupling alignment can be easily compensated by attaining the same spectral broadening with the same input laser power. Finally, the compact size allows the whole microscope to be integrated into a movable cart for clinical study. These advantages make this new SLAM system a powerful and user-friendly imaging platform.

**Funding**

This work was supported in part by a grant from the National Institutes of Health, U.S. Department of Health and Human Services (R01 CA241618).

**Acknowledgement**

The authors thank Darold Spillman for his administrative and technical support, Eric Chaney for help with the mouse surgery, and Janet E. Sorrells and Rishyashring R. Iyer for help with the solution preparation.